# What is an Observatory?
# The Crucial Role of Such Organizations for
# Maximizing the Science Return from Astronomy Research Facilities

**Garth Illingworth**

March 30 2013
Updated February 10 2017

## Abstract

*Observatories, Institutes and Centers that support major astronomical facilities have, for decades, played a key role in maximizing the scientific return from their facilities and telescopes. It is crucial to realize that an "observatory" is more than simply a telescope or ensemble of telescopes that provide observing opportunities for astronomers. Observatories are at the heart of an institutional capability to do things beyond what is possible for astronomers themselves to do as individuals. Observatories organize and mobilize a range of multidisciplinary skills for achieving a coherent and sustainable capability that is central to modern observational astrophysics. Yet many of the activities and roles carried out by the "observatory" organization in fulfillment of their mission are not fully recognized by policymakers or senior managers, and often not fully understood even within the astronomy community. As we assess the future opportunities for new facilities in a challenging budget environment it is important to be cognizant of what observatories do and the roles they play.*

## UC Context

UC astronomy represents a world-class capability that is built on long-term facilities (its telescopes) with operational lifetimes that cover many decades. The UC Observatories (UCO) has enabled UC and its facilities to achieve world-class status and recognition, nationally and internationally. A recent UC-wide Astronomy Task Force (ATF) defined a series of priorities for UC astronomy. These priorities, and the importance and accomplishments of UCO, were then endorsed and highlighted by a highly experienced international committee of senior astronomers and managers (the External Review Committee). Responding to the recommended changes in priorities will require changes at UCO in staffing and capabilities. Such staffing and capability changes are a valuable opportunity to optimize efficiency and productivity, and to refocus the observatory on future needs. Yet the impact on long-term employees and the UC budget situation requires UCO and UC astronomy to give careful consideration of how to respond to the new priorities. The changes need to be made thoughtfully, utilizing an understanding of what makes an observatory function well (as UCO is recognized to have done), and done with care to maximize the future potential while minimizing the damage. In making the needed changes it is crucial to understand what an "observatory" does and why a centralized operation plays such key roles in (1) the utilization of astronomical telescopes, (2) efficient operation of those facilities, and (3) upgrades that keep the telescopes at the cutting edge of scientific research. This "white paper" lays out some background regarding the roles played by observatories in maximizing the scientific productivity of long-lived facilities and telescopes, followed by a more focused discussion of the role played by UCO. It concludes with some of the challenges facing UCO. The discussion is laid out as follows:

*(1) The budget challenge and evolving priorities.*

*(2) The priorities.*

*(3) What do observatories do?*

*(4) Has UCO been an observatory in the commonly accepted sense?*

*(5) UCO and the future of UC astronomy.*

*(6) Does UCO actually have a future?*

*(7) The appendix: the author's experience summary, the Astronomy Task Force and External Review Committee executive summaries.*

## Foreword

The attached "white paper" provides some thoughts regarding Observatories and the role they play in maximizing the return on the long-term investment in facilities that is inherent in ground-based astronomy projects. While much of the discussion is focused around the UC Observatories, the opportunities, issues and challenges are generically applicable for Observatories as a class, though, of course, there will be differences in detail.

I hope this proves useful for thinking about the future of astronomy at UC. The loss of an identifiable observatory that is at the heart of UC will lead to fundamental and deep-seated changes. The synergistic effect of having a UC Observatory working cooperatively with a set of strong astronomy departments has made UC astronomy arguably the most effective and visible university astronomy program in the world. UC astronomy without an observatory will be greatly weakened.

UCO has brought great visibility to the UC astronomy program, particularly amongst federal funding organizations. It is common for university groups to establish an "institute" or a "center", or in the case of astronomy, an "observatory". Such organizations have clearly defined and easily recognizable goals (e.g., like Scripps at UCSD or SLAC at Stanford). The visibility that these "brands" bring to universities is of great value in fund-raising, both government and private, and in enabling simple name-recognition identification as a player on the national and international stage. Having a "cross-UC" astronomy "brand" has been very valuable. *It would be very unwise to throw away decades of "UCO" brand recognition, and over a century of "Lick" brand recognition.* Lick was a pioneering and highly visible example of a large private donation that led to a world-leading astronomical facility, just as the private donation to the twin Keck telescopes did 100 years later, and the donation from Gordon Moore is poised to do for TMT.

The focus of the current discussion on eliminating or dramatically cutting the UCO faculty and on building "instrumentation" as the only important activity of an observatory is a mistake. It demonstrates a serious lack of understanding of the central role that a group of managers (project managers; system engineers), engineers (software and hardware) and scientists working as a committed and focused team play in today's large, high-technology projects. Doing away with such a team co-located to enhance interactions would lead to a huge loss of capability, and would damage the future of UC astronomy. A core, co-located group provides both efficiency and administrative responsiveness. A core group that maximizes opportunities for facilities, while working to enhance the capabilities of groups at other campuses (like the UCLA Infrared Laboratory), provides depth and longevity to the facility side of the overall UC astronomy program.

Nationally, the visibility of UC astronomy will decrease if UCO goes away. Our collective ability to raise funds for instruments and facilities will be weakened. UCO has always been accorded a great deal of respect within the Astronomy Division at the National Science Foundation (NSF) for its accomplishments and experience as an observatory. UC astronomy has benefited from this visibility and credibility. It has been important for Keck, and it will be important for TMT.

It is crucial that the role that an observatory plays receive careful and objective consideration, especially in an era of tight budgets, when optimizing the overall UC astronomy program becomes even more challenging if the damage from funding cuts is to be minimized.

The following "white paper discussion" assesses the roles that observatories play, and highlights the activities that they perform to carry out their mission. The review then discusses the role that UCO has played vis-à-vis Lick, Keck, and TMT, as well as its role as a key entity for UC astronomy.

## (1) The budget challenge and evolving priorities.

UC has been facing significant budget problems and it is clear that a thoughtful, thorough and careful evaluation is needed of the system-wide aspects of UC astronomy and our facilities. The strength of our campus programs and of our astronomy faculty is well known and very visible worldwide. What has made UC stand out is that UC has had an observatory at its core. UCO played a central and key role in initiating efforts that led to Keck, in making Keck the powerful facility that it has been for the last 1.5 decades, and in initiating the Thirty Meter Telescope (TMT) concept and program. The two strong instrument labs that are part of UCO, one at UCLA and the other at UCSC, have routinely delivered astronomical instruments for UC telescopes that are the best in the world. The synergy between a centralized observatory and a distributed UC-wide research capability has been extremely effective. If there are doubts about this, one should reflect on the statements of the recent External Review Committee (see their Executive Summary in the appendix). This was one of the most experienced, competent and strong committees that I have seen undertaking a review (and I have initiated or participated in many such committees for observatories). The Review Committee's statements about UCO were unusually positive for such an experienced committee.

Regardless, the reality is that times change, and organizations need to evolve to respond to changing circumstances, as the Review Committee itself acknowledged by endorsing a set of priorities from the Astronomy Task Force (ATF). Times of fiscal stress are a challenge, but are also an opportunity to reinvent an organization to do better within the available resources. Reinvention is a valuable process, when it is carried out in a constructive, thoughtful, transparent and involving way. It can be worrying to those trying to defend the status quo because they might lose some capability that they prize, but looking beyond these parochial interests is crucial. The opportunity to refocus the organization to mesh with new priorities comes with pain, but it also can lead to a stronger organization. When this need for reinvention coincides with budget issues the result will be cuts and a reduction in the capabilities. Again, however, when the changes are made with care the result can be a stronger organization (as long as the cuts are reasonable – there are levels of cuts that will destroy any organization regardless of its strengths).

One guide to thinking about the future of UCO is to ask whether the future UCO actually constitutes a recognizable entity that would be reviewable in 5-10 years by a committee like the last External Review. Some models of distributed and fluctuating groups occasionally doing instruments may well not even constitute an entity that makes sense to consider as an "observatory", and hence to review. There may well be a model for UC astronomy that doesn't have a distinct entity called UCO. But does this model make sense? Let us come back and address this question once we have considered what observatories do, since we have many successful examples.

## (2) The UC priorities.

The first step needed for a careful reevaluation of any organization is to get concurrence on priorities from the stakeholders and customers. The Astronomy Task Force (ATF) process, with the endorsement of the ATF priorities provided by the very experienced External Review Committee, gave a framework of priorities within which this reevaluation should take place (see the attached Executive Summary). The ATF reported the following priorities from a community-wide survey, along with the percentage of support for the different facilities (see the ATF executive summary and report). This extract is taken directly from the ATF Executive Summary:

*The survey of the UC A&A community clearly identifies the following prioritized ranking of facilities for UC system-wide investment (and associated percentage support):*

1. *The Thirty Meter Telescope (TMT) Project (90%)*
2. *Keck Observatory (89%)*
3. *UC Instrumentation Labs (70%)*
4. *Lick Observatory (40%)*

*Other proposed investments included LSST, the Large Synoptic Survey Telescope (20%), a system-wide facility for astrophysical computations (16%), and a radio astronomy facility (10%).*

The highest priorities for UC Astronomy and Astrophysics community were clearly TMT and Keck, followed closely by the UC Instrumentation Labs.

Thinking about how to optimize the returns within a smaller budget is a challenge, but it is also a very rewarding activity when approached objectively with the goal of maximizing our scientific capabilities and our ability to provide scientific data. A key to this next step is the Strategic Planning Committee (SPC). This UC astronomy-wide committee is now in place; it needs to develop a more detailed strategic plan and vision for UCO and UC astronomy for the future. This will be hard given the current pressures on UC astronomy, but it should still be carried out (and it would be good to carry out the SPC process expeditiously). Along with the input from the other groups (ATF; External Review), the SPC input will be a valuable resource for the UCO Board, for the UCO Director and faculty, and for the UCO Advisory Committee (UCOAC).

This planning process for astronomy naturally has longer-term implications beyond the usual 5 years because of the lifetime of our facilities. Lifetimes of 30-50 years are not unusual for the basic astronomy facility (the telescope). The reason these facilities last so long at the scientific forefront is that we can, and do, rejuvenate their capabilities on decade or less timescales as we improve the instrumentation, operations, and facility performance. Adaptive Optics is an example of a capability that will completely revolutionize the performance of our facilities. Astronomers are constantly reinventing and enhancing their facilities to provide state-of-the-art capability at a fraction of the facility cost. This makes astronomy research programs quite cost-effective in the long-term. A new report, the *Cost of Astronomy*, is near to completion. This report quantifies these costs for astronomy vs. laboratory science in UC (and shows that astronomy is typical of other sciences – see § 6).

While it is clear that we need to evolve and to reevaluate the UCO mission and capabilities, it would be valuable to do so in the context of what observatories do, and why the model for successful observatories is remarkably consistent world-wide. This review is based on my 35 years of working at observatories, doing large projects, overseeing observatory management and oversight and involvement with major projects (see the appendix).

## (3) What do observatories do?

I will comment first on observatories in general, since there is much commonality worldwide in the roles and structure of observatories that have developed over the last 30-40 years. There are some differences for a UC observatory that I will then comment on.

What is an observatory? Observatories have diverse roles. Obviously there are differences in detail between those that deal with space missions and those that deal with ground-based facilities. And there are differences among different observatories. But the broad structure, the type of staff and the objectives are surprisingly similar across observatories. Aspects of observatory activities vary with time (e.g., new facilities and instruments and upgrades), while others are more ongoing (operations and science support).

*Activities:* Activities at observatories are both numerous and diverse:

1. Facility operation and management (requiring, e.g., operations personnel, support and maintenance capabilities, facility and staff management, business office support, etc.);
2. Science support (e.g., time allocation, archive systems, analysis systems, observing support);
3. Support for advisory structures and committees to ensure stakeholder and customer involvement in optimizing the observatory's activities;
4. Enhancing facility science productivity (e.g., operations improvements);
5. Problem recovery (e.g., fixing instrument failures, optics problems, software upgrades);
6. Facility and capability upgrades (hardware and software – e.g., optics, coatings, telescope and instrument software);
7. New facility capabilities (e.g., Adaptive Optics (AO), upgraded optical components, detectors and systems);
8. Instrument improvements and enhanced capabilities;
9. Design and development of new instrument concepts, instrument R&D;

10. Development of cost and schedule plans for instruments and upgrades;

11. Establishing reviews and review material for the potential new instruments/upgrades;

12. Proposing, fund raising (public and private), and ensuring cost matching for new instruments and facilities;

13. Public outreach and responding to public interest in astronomy;

14. Education and training in those aspects unique to an observatory;

15. Construction of new instruments, testing and integration of instruments and upgrades;

16. Conceptual development of potential new facilities (e.g., new telescopes);

17. R&D for critical hardware/software systems for new facilities;

18. Development and demonstration of cost and schedule plans, including demonstration to review committees of the reality and viability of the program;

19. Overseeing and management of the development phase of the new facility;

20. Management of the construction of the facility;

21. Facility acceptance and operational testing and problem fixes.

Clearly there are many tasks, activities and responsibilities that are required of an observatory – no doubt there are others that I have not listed!

*People:* Activities like those listed above are carried out by observatory staff, both internally as projects, and externally by managing contracts and providing support for external groups and companies. Observatories do these activities by utilizing the experience of the staff, the breadth of their skills (engineering, management, science, technical), and the synergism that arises by focusing on common objectives with experienced people possessing diverse skills and backgrounds.

A crucial element of a successful observatory is the quality and diversity of its staff. Projects, large and not-so-large, telescopes or instruments, be they in space or on the ground, work well only when there is a strong project management team (Project Manager, Systems Engineer) with strong engineering/software support, and a dedicated and focused science group (astronomers and physicists). The size of these groups depends on the scale of the project, but the core elements – project management, engineering/software, science – remain critical to the success of any major project. Each of the three core groups brings crucial skills to the table. The synergy that operates within a management/engineering/science team plays a central role in the ability of an organization to complete a technically demanding project consistent with the budget and schedule.

Since the projects we are discussing are science projects, the scientists involved have a central role to play and have key responsibilities. They develop the science objectives and requirements, they work to ensure that the concepts and their capabilities are consistent with those objectives and requirements, they involve students and postdocs as part of their training responsibilities, and they continue these oversight roles during the project, utilizing their experience to guide the choices and decisions so as to minimize the impact of technical issues and challenges on the final scientific performance. The scientists also play a central role in fundraising and public outreach, especially when the scientific results are forthcoming from the new instrument or facility.

***The roles of observatories in a historical context:*** The multiple roles that an observatory plays have actually changed little over the decades, evolving insofar as technology has evolved. The core aspects of observatories were already apparent in the 1970's. It may seem strange to go back this far, but this was a crucial period that set the stage for all our space observatories. Ground-based observatories at that time were, with varying degrees of success, well-established entities (National Optical Astronomy Observatories, National Radio Astronomy Observatory, European Southern Observatory, Lick, Carnegie, Palomar, Steward, McDonald, Smithsonian Astrophysical Observatory, etc. – and I am surely missing others and have not listed most of the international observatories).

The discussion about whether to use this model for space observatories, particularly for Hubble (at that time "Space Telescope"), was intense. NASA was very unsure about the path it wanted to follow, but the relatively new Office of Science asked for a National Academy of Sciences (NAS) study and report. The result was a very thoughtful NAS study – the Hornig report. This study, *Institutional Arrangements for the Space Telescope*, was carried out in 1976 by the National Academy of Sciences. The committee was a diverse and experienced group chaired by Donald Hornig, and it dealt with Institutes/Observatories and their roles. The Hornig report recommended a Space Telescope Institute (which became Space Telescope Science Institute – STScI), and did so with a list of 27 recommendations, many of which were peculiar to NASA and the quite new concept of a space observatory. Many, however, are relevant in today's world for all observatories. The appropriateness of the Hornig report's recommendations was verified by another (but smaller) National Academy study in the mid-1980s *Institutional Arrangements for the Space Telescope – A Mid-Term Review*.

One recommendation, taken directly from the 1976 Hornig study, is noteworthy because of the current situation (and, in particular, note the statement regarding "independence"):

*6. We recommend that the policies of the STSI be set by a policy board of about ten people representing the public interest, as well as the astronomical community and the broader scientific community. The quality and independence of the policy board is essential to the success of this enterprise.*

What was very clear in both the original Hornig report, and the mid-term review in 1985, was the emphasis on a strong scientific staff as a key aspect of the space telescope observatory. For example, the 1985 report stated:

*2. The ST/ScI must continue to attract and maintain a scientific staff of the highest quality. This requires a vigorous in-house program of scientific research.*

This focus, on the role of the scientific staff and the need for a high quality scientific staff, arises often in discussions about observatories/centers/institutes. It is noteworthy that essentially all observatories have settled on a 50-50 scientific staff (50% research, 50% observatory duties) as appropriate for the dual goal of having a scientific staff who are committed to the observatory, but who are scientific peers of those in the observatory's user community and who can command respect within the funding organizations and agencies. The key here is that there is a faculty comparable in quality to the faculty in the organization's "customer" base, but who have a personal and professional commitment to the observatory.

In short, my direct experience with the Hubble Space Telescope (HST), Keck development and construction, Next Generation Space Telescope (NGST – now the James Webb Space Telescope JWST) development, TMT development, instruments for HST (Advanced Camera for Surveys) and Keck, and several decades of overseeing instrument and facility development on committees and boards has emphasized again and again the need for a strong, effective experienced **team** (in the clearest sense of the word) of managers, engineers/technical staff and highly-respected scientists working closely and synergistically together with mutual respect and understanding of their respective roles and contributions.

The bottom line summary for observatories is four-fold:

1. Observatories carry out a wide range of activities; no single activity is dominant;
2. The synergy between management, scientists and engineering/technical staff is key to remaining at the cutting edge;
3. Being world-class requires a team of committed and experienced people;
4. It is vital to have a core scientific staff/faculty (the 50-50 staff) who are widely respected for their research, for their understanding of technical issues and management, and for their long-term commitment to the observatory.

## (4) Has UCO been an observatory in the commonly accepted sense?

UCO has clearly been an observatory that matches with the four aspects immediately above. The External Review Committee judged UCO in the world-wide context of observatories and stated:

*1. By all criteria the performance of UCO as an organization that supports and advances observational astronomy within the entire UC system has been excellent. Objective evidence for this excellence includes:     (see the sub-bullets in the full executive summary in the appendix)*

It is very clear from the context and the tenor of the External Review Committee's report that they judged UCO to be a very capable and successful observatory.

*New telescopes:*  The achievements of UCO over the last couple of decades have been remarkable. Both the Keck telescopes and the TMT owe their existence to efforts at UCO to make sure that UC astronomy had access to forefront facilities.

The TMT project would not exist today if it were not for the efforts of a number of key people at UCO. Jerry Nelson, Terry Mast, Joe Miller, Mike Bolte, and several of the engineering staff (and others I am sure) were central to both developing the concept and providing initial support to ensure that TMT could get from the realm of being just a vision to being a potentially viable project. As with many start-ups, a vision is but the first step. The really challenging phase, as venture capitalists know, is turning the vision into reality.

The same was true for what ultimately became Keck. Lick Observatory (as it was called prior to the change to UCO in the 1990s) faculty and its leaders realized in the late 1970s and early 1980s that UC astronomy had to develop new telescopes to ensure that UC astronomy remained at the cutting edge. A large telescope on Junipero Serra was an early consideration. The approach of using a single large mirror was championed at Lick. This ultimately was replaced by the remarkable segmented-mirror concept that was conceptualized and then demonstrated by Jerry Nelson and others at Lawrence Berkeley National Laboratory (LBL). But, despite the disappointment, particularly on the part of some faculty, Lick Observatory did not walk away from the vision of a new telescope for UC astronomy. Lick Observatory and its faculty went on to embrace this new technology and poured effort and resources into supporting the new segmented-mirror telescope.

*Keck first-light instrument management:* UCO people played major roles in making sure that the needs and aspirations of the UC astronomy community were realized through Keck. The efforts on the Keck Board by Bob Kraft, Joe Miller and by senior administrators at UCOP, along with their Caltech counterparts, provided the management and political support to move Keck forward, working with the Keck Project Manager.

The same was true of the Keck Science Steering Committee (the SSC). Those not familiar with the Keck project during construction may not realize that the Keck Project Manager, Jerry Smith, decided that he would not manage the instrument development for either of the Keck telescopes. The oversight of instrument development for Keck was to be done by the SSC. The budget for all instruments was assigned to the SSC, and it was expected that the project management, monitoring of funding and oversight would be done 100% by the SSC. This was not normal procedure elsewhere, nor was it NASA practice. This is not an ideal approach, since the SSC scientists lacked the experience and project management background, but UCO and Palomar scientists accepted the situation, embraced the challenge and, as a result of a great deal of effort and commitment of time carried out the required oversight.

While the individual instruments had lead engineers and scientists, essentially none had experienced project management. Normally the overall facility project would provide such support. Since the Keck project did not do this, and since the chair of a committee does the vast majority of the work, the outcome was that the Caltech and UC co-chairs of the SSC essentially became project/program managers for a decade through the 1990s (Sandra Faber, then Garth Illingworth on the UC side from UCO, and Wal Sargent, Tom Soifer and Chuck Steidel for Caltech; Mike Bolte then continued the UC leadership on the SSC at the end of the 1990s). Being Chair/instrument project manager for Keck was a very demanding and highly time-consuming activity. This really only worked because the scientists from the respective observatories, and particularly UCO, had enough background to step into the required roles, and could commit the time required year-round.

*Keck instruments:* There were numerous examples of activities at UCO that took place in the 1990s that made the Keck telescopes the productive facilities that they are today. UCO at UCSC and at UCLA played a central role in the first major set of Keck instruments. Steve Vogt and his superb UCO team put in a remarkable effort on the High Resolution Echelle Spectrometer (HIRES) that led to one of the most professional deliveries ever of an instrument to Keck. The Low Resolution Imaging Spectrometer (LRIS) was primarily a Caltech effort, but would not have been realized without its UCO contributions. Joe Miller played a key role in the LRIS concept development. Harlan Epps designed and oversaw the camera development. The UCLA Infrared Laboratory group led by Ian McLean built and delivered the very challenging Near Infrared Spectrometer (NIRSPEC) instrument as part of the initial complement of Keck instruments. Joe Miller and Mike Bolte led the Echellette Spectrograph and Imager (ESI). The DEep Imaging Multi-Object Spectrograph (DEIMOS) was built at UCO and led by Sandra Faber.

More recently, the UCLA Infrared Laboratory has built new IR user instruments for Keck, the OH-Suppressing Infra-Red Imaging Spectrograph (OSIRIS) and the Multi-Object Spectrometer for Infra-Red Exploration (MOSFIRE), under the leadership of Ian McLean and James Larkin (MOSFIRE also involved Chuck Steidel from Caltech). These instruments take advantage of the major gains in IR detector technology since the first generation IR instruments at Keck. The recently delivered MOSFIRE is the most powerful IR spectrograph in the world today.

The DEep Imaging Multi-Object Spectrograph (DEIMOS) provides a striking example of synergistic activities by scientists and engineers that are necessary for carrying out a successful instrument project. A short summary of the history of how DEIMOS came about would be a relevant example. I am sure there are many other such stories, but this is an example that I know about directly.

***DEIMOS:*** This major instrument was an excellent example of how faculty and engineers needed to work together to enable the project, and then ensure its success. DEIMOS started in somewhat unusual circumstances. David Koo, after some initial contacts and discussions with the leadership of the new Center for Particle Astrophysics (CfPA) at Berkeley in 1990, realized that the Center might well be interested in supporting the development of a new capability at Keck that could accomplish some of the Center's science goals (those related to cosmology and dark matter). David then organized further discussions with the Center leadership (Bernard Sadoulet) and Center members, and asked me to become involved because of my background in instruments and facilities, particularly for Keck. They were interested and so I developed the concept of a wide-field, double-barreled, optical multi-object spectrograph. Harland Epps and others at UCO played a key role in fleshing out the DEIMOS concept that we presented to the Center, and ultimately in an NSF proposal, and to the Keck SSC and the Keck Board.

The Center for Particle Astrophysics provided some seed support. While modest, it was, nonetheless, crucial as a way of demonstrating that our DEIMOS concept was of wider interest, and helped considerably when we wrote a proposal to the NSF. Sandra Faber was very interested in the project too and was adding her considerable energy and talents. I PI'd the NSF proposal which was then funded at the level of $1.79M by the NSF for us to continue to develop the concept and to verify some of the key technologies. This was a striking example of the synergies that need to happen to get facilities and instruments started, accepted, built and commissioned. Multiple people with different interests and skills must be involved (the initial phases alone involved five science faculty, David Koo, Garth Illingworth, Harland Epps, Sandra Faber, Joe Miller plus several engineers, along with budget and schedule analysis professionals). Having all these people in one organization with a great interest in developing a forefront capability was key to the development of DEIMOS.

At the core of all this effort for the faculty leading the program was the science goal: to be able to carry out the Deep Extragalactic Evolutionary Probe (DEEP) survey. This gave the critical focus, energy and motivation to having DEIMOS be as powerful and as efficient as possible. [As an aside, I was doing the HST ACS camera also at that time with Holland Ford, and so I took over being Chair of the Keck SSC and Sandy became PI of DEIMOS – thereby removing a key conflict of interest. The PI of an instrument should not also be the Keck instrument program manager – in this case the Chair of the SSC.]

One other aspect of the DEep Imaging Multi-Object Spectrograph (DEIMOS) story was an international effort involving a possible collaboration with the Japanese to obtain private funding for the second side (second "barrel") of DEIMOS (DEIMOS was conceived as a double-sided instrument, but eventually only one side could be built because of budget constraints). Sandra Faber, Raja Guhathakurta, David Koo, Garth Illingworth and Harland Epps had managed to interest the President of Fujifilm in having Fuji support the second side of DEIMOS. A very successful visit to UCSC by the president of the company was being followed by discussions regarding funding when a number of unfortunate events coincided amongst the many players involved (health problems; the Fuji president being replaced; a change in the economic situation for Fuji). The whole plan collapsed. This was a substantial effort involving many people, and while it was ultimately not successful, it is an example of how UCO faculty have also worked hard as a team to do fundraising, like other faculty across the system.

DEIMOS ultimately took a dedicated effort of over a decade to develop, build and commission (it was commissioned by late 2002). Having a faculty group willing to spend this much time and energy was key to the success of the DEIMOS project. The Deep Extragalactic Evolutionary Probe (DEEP) survey began in 2003 and initially continued for 5 years. During this time DEEP evolved to DEEP2 and became a major, UC-wide collaboration, particularly with UC Berkeley, and also adding national and international team members. The DEEP projects also had a strong education and training component by involving large numbers of graduate students and postdocs.

The key lessons to be learnt from this example (and it is not unique for projects of this scale) is the need for a long-term commitment by a scientifically-motivated faculty team, and the need for a highly capable, diverse and experienced team of faculty and engineers and project managers.

*Keck problem-solving:* It was also during this period (the 1990s) that Jerry Nelson and Terry Mast jumped at the opportunity to move to UCO. They recognized that an observatory like UCO was their natural home within UC. Their talents, skills and interests were fully recognized and supported in the observatory environment. Furthermore, the observatory environment enabled them to work effectively and efficiently with a capable group of engineers, technicians, research scientists and software professionals to ensure that the Keck telescopes worked superbly, consistent with their potential performance.

The Keck Observatory has, as was planned, utilized the intellectual and other resources of UCO and Palomar to help fix problems, to optimize the performance of the two telescopes, and to make their operation as efficient as possible. Naturally as the bigger observatory with a much larger range of people, UCO was on the forefront of most of the efforts to deal with problems and improvements at Keck that were beyond the capabilities of the local staff in Hawaii. Issues with actuators, coatings and most recently segment repairs have required support from UCO people to find solutions and develop fixes.

In addition, UCO staff developed much of the software for distributed control of instruments for Keck, as well as all of the remote observing systems being used to allow mainland observing on Keck by UC faculty and others.

*Keck facility upgrades and AO:* While UCO played a key role in leading a number of Keck instruments, what has not been so clearly recognized is the effort that goes into working on updates to facilities at Keck and on upgrades to instruments. In recent years, UCO/UCSC efforts have been directed primarily towards upgrades. This is a natural outcome of the initial efforts that focused more on building major *optical* instruments for Keck because of the maturity of optical detectors relative to IR detectors in the 1990s. The UCSC optical detector lab and its developments have played a substantial role in numerous instruments and upgrades. The detectors of the Low Resolution Imaging Spectrometer (LRIS) were upgraded at UCO. UCO also did the High Resolution Echelle Spectrometer (HIRES) focal plane upgrade in 2004.

During the last decade, when the focus was on optical instrument upgrades at UCO, the UCLA Infrared Laboratory, built, as noted above, new IR instruments such as the OH-Suppressing Infra-Red Imaging Spectrograph (OSIRIS) and Multi-Object Spectrometer for Infra-Red Exploration (MOSFIRE). Interestingly, with the increasing maturity of IR detectors, the time has come for upgrades to become the focus of attention for the older IR instruments like the Near

Infrared Spectrometer (NIRSPEC) and even OSIRIS, and so the UCLA lab is about to embark on a series of upgrades

The Keck Atmospheric Dispersion Corrector (ADC) was built at UCO. It was led initially by David Koo and then completed by Drew Phillips and Joe Miller. Ongoing issues with software at Keck were very often resolved with the crucial help of Bob Kibrick. Hardware and software issues with the telescopes often required UCO faculty and researchers like Jerry Nelson and Terry Mast and others to identify solutions and fixes.

The Center for Adaptive Optics and the Moore Foundation Laboratory for AO are very successful activities at UCSC that are closely linked to AO developments at Keck and will play a central role in the scientific utilization of the TMT.  This has involved many UCSC people (Jerry Nelson, Claire Max, Don Gavel and many others).  The AO Center has carried out a highly regarded training program, has involved numerous graduate students and postdocs in its programs and has run a very effective outreach program as well.

*Mt Hamilton:*  In addition to all these activities UCO has operated Mt Hamilton on increasingly limited resources.  Mt Hamilton is not a large modern observatory, but it does have unique capabilities that are scientifically important and scientifically productive. It also provides opportunities for observing and training that are very appropriate for a university-based facility, and complements the Keck telescopes. UCO has also supported facilities there that are more PI-like. UCO has spent over $2M of its resources on the Automated Planet Finder (APF), while Steve Vogt has invested a huge share of his personal time and effort into trying to complete the APF project.

There is another aspect to having UCO operate Lick Observatory that does not get much attention. Lick Observatory is UCO's primary source of experience for operating telescopes. While UCO staff are involved in many issues related to the operation of Keck, UCO does not directly operate Keck.  Similarly, it will not operate TMT.  There is a mix of technical and software skills that are required for operations that are valuable to have in an organization that is dedicated to the fabrication of upgrades and instruments for operational facilities, especially when that organization will be asked often to help in problem-solving at Keck and TMT (and any other observatory of which UC will be a part).  But operating any facilities, even small facilities, comes at a cost. The need for people, technical and support, business office support and maintenance, in a modern day environment with safety constraints and work-rules necessarily makes operation, even of a small facility, quite costly.

Budget issues and the priorities endorsed by the Astronomy Task Force and the External Review Committee may preclude continuing operation of Lick, but in my view it would be unfortunate, both because of the potential loss of Lick facilities for UC students, postdocs and faculty, and because of the loss of operations experience that Lick provides for UCO.

*Instrument development:*  Instrument development is often discussed, and it is a key part of what observatories do, but, as the long list of activities above discuss, it is by no means the only aspect of the activities of an observatory – even though some of the recent discussion within UC would suggest that it is the only one that is important.  This is, unfortunately, rather naïve and reflects a limited view of what observatories do, and what has made UCO and UC astronomy world-class.

Development of instruments directly by UCO in Santa Cruz has been a key activity for UCO. While instrument development at UCO needs to remain a key activity both for TMT and Keck, it is not the only location where instruments could, or should, be built. UCO needs to enhance the ability of groups outside UCSC to build instruments. The UCLA group has also been very successful in its instrument developments and is an excellent type-model that can be utilized for the future.

This "external" model has a long history that can be used to help understand both the benefits and the challenges of developing instruments in a distributed way. For example, the European Southern Observatory (ESO) has, for a long time, chosen to do its major instruments by external groups. This was initially in significant part for political reasons (the need to spread resources amongst the member countries), but it is now recognized that the external approach has enabled instruments that utilize technology, capabilities and experience are not readily available in-house in Garching at ESO HQ. The UCO experience with UCLA has likewise been both positive and valuable in terms of broadening the type of instruments that can be built. Gemini also takes an external approach. The National Optical Astronomy Observatory has also realized that there are benefits to having instruments built by others.

It is a mistake, however, to think that an instrument for huge telescopes like the Keck 10-m and the TMT 30-m telescopes can just be handled by a simple contract with an external group. ESO's long experience with the "external" approach provides an example of the level of involvement that is needed to make sure that the instruments perform well. Considerable ESO resources, personnel and expertise at ESO headquarters are used to ensure that the instruments meet the required standards and perform as required.

To institute an "external" model of instrument development requires that everybody involved be very realistic about the challenges and costs of doing a modern instrument for a 10-30 m telescope. Such instruments require a level of professional capabilities (management, engineering) and ways of doing business (adherence to requirements, careful attention to cost and schedule, monitoring of progress, etc.) that are not easily replicated, or readily available in an environment where instruments are done only occasionally. Nonetheless, opening up the opportunities for instrument development should be done to enhance the breadth of deliverables that can be accomplished under the umbrella of an observatory. But doing this does not obviate the need for a centralized operation with a broad skill mix of people to provide oversight, support, and expertise in the many areas that are impractical for an external group to maintain and support long-term.

In the case of Keck or TMT, many of these functions can and will be provided by the Keck and TMT organizations, but it would be a serious mistake for UC astronomy to devolve all of that responsibility to them. Experience has shown that Keck Observatory, for example, benefits from its relationship with UCO personnel and experience in a number of areas including instrument development. Furthermore, if the UC astronomy community wants a particular instrument, particularly for TMT, getting what the UC community wants with the performance we demand will require us to take a major role in developing that instrument. That ability will be compromised if we do not have the required mix of experienced people within UCO and its campus groups.

The bottom line, as experience has shown for instrument development in many environments, both space and ground, is that an observatory plays a key and central role. The synergy between smaller external groups with particular expertise and the core observatory with the experience and continuity of capability and personnel is what makes great instruments. UCO has demonstrated this well with its involvement in instrumentation from both Caltech and from UCLA.

***Summary:*** I think it is clear that UCO is, as the External Review Committee recognized, a true "observatory", with a broad base of skills and a mix of people that have allowed substantial progress in many areas. The mix of scientists (astronomers and physicists, faculty and researchers) and engineers (software and hardware) at UCO has worked well in the past, providing the right synergies for the activities at those times, but the mix now needs to evolve for the future given the changing priorities. While it would be good to have UCO able to evolve to what is needed for the future, there are serious challenges ahead, and not only with budget. These challenges are discussed in § 6. The viability of UCO as an observatory is at stake because of these challenges. Before the challenges are addressed I will discuss those aspects of UC astronomy that would be seriously impacted if UCO was eliminated.

## (5) UCO and the future of UC astronomy.

The budget pressures are a challenge, but can be used to make changes that position an organization for the future. Organizations can come out of such exercises stronger and more relevant to the issues of the future. However, this requires:

1. A rational process of review and assessment by those funding the program with the goal of optimizing the program;

2. An understanding of how observatories work and what makes them successful;

3. A willingness to step beyond parochial self-interest to find compromise solutions that are optimal for the overall program;

4. A willingness to work with observatory people who are trying to develop options for the future.

As I tried to indicate in the previous sections, observatories carry out a wide range of activities, and UCO has not been different, except in detail, from many/most other observatories.

While a lot of options are being explored, the discussion seems to often to focus on the UCO faculty, with little or no consideration given to the key role that the research astronomers and physicists play, or the engineers, or the technical people, or even the business people who deal with high-technology contracts and the complex grants and funding streams that are quite different from those dealt with by campus business offices. In the extreme, the view seems to have arisen in some quarters that a few faculty with some "observatory roles" plus a few other faculty scattered around the system can do it all and that a centralized observatory is not needed. This is naïve and it is not been the experience for successful observatories anywhere else.

A related view is that doing instruments is the only significant aspect of what an observatory does. Even with a separate Keck Observatory, and even if the operation of Lick Observatory declines in scope in future, a UC observatory will still do far more than "instruments", as should be obvious from the above discussion. If all "UCO" does in the future is "instruments", particularly in a distributed approach, UC will not have an observatory.

Several key capabilities would disappear if UCO as we know it today disappears:

***(a) future UC astronomy facilities.*** Any future UC astronomy facilities (new telescopes) are highly unlikely to happen without UCO. TMT is not the last telescope that UC astronomy should aspire to building or to share in the construction of. TMT will be immensely powerful, and even a modest share (<15%) would provide remarkable opportunities, but a modern (smaller) telescope that provides a large share for UC scientists and students would open up different opportunities. In the unfortunate situation that TMT does not go ahead, the pressure for a new facility will rapidly grow even stronger. In either case, conceptualizing, developing and building partnerships takes time that invariably falls on the shoulders of UCO, and particularly on the shoulders of its faculty. The combination of faculty focused on long-term future facilities working with experienced engineers and managers is what generates new telescopes and capabilities.

*(b) UC as a strong partner in TMT.* TMT, like Keck, will just not happen. TMT has a larger partnership, with many very good scientists, but the experience base with major facilities is not deep. The value of very experienced and committed scientific involvement will be one key component of its success. The sense that I get at times is that there are those who think that a distributed UCO faculty interested in TMT would be enough to ensure UC astronomy a significant voice in TMT. This is just not true. Again, it is the synergy aspect where dedicated effort by managers, engineers and scientists is what makes one a player. Scientists are notorious amongst project managers for thinking that they know far more than they do! Strength comes from the hard-won respect that comes to a group that is recognized as "knowing what they are talking about." Developing respect from the project manager and project team takes time, but when it is achieved, that respect leads to opportunities for leadership in deciding future priorities and instrument choices in a project.

*(c) support for Keck Observatory.* The Keck Observatory will continue to need help on facility upgrades, problem resolution and fixes that require people and technology expertise and experience that is not available in Hawaii. Since a number of senior people at UCO are retiring and their extensive experience base is disappearing, it would we wise to plan for hiring people who could continue UCO's role in support of Keck. Instrument experience is part of what UCO is losing. It is striking that the majority of the major facility instruments that have been delivered to Keck, and all the facility instruments for Keck on the UC side have so far been from the two UCO instrument groups, at UCLA and at UCSC. UCO has an excellent group at UCLA and that should be enhanced, but, as the Infrared Laboratory Director Ian McLean has noted, they benefit from the support and interactions with a larger group at UCO. A weak or non-existent UCO would lead to weaker and less capable groups elsewhere within UC.

*(d) delivering instruments optimized for UC astronomer needs.* The European Southern Observatory (ESO) does essentially all of its instruments as "community deliverables". Gemini is similar. In these, and other observatories, groups outside the observatory are building the instruments. What is striking in ESO's case (as an example for which I have more background) is the degree to which the instruments are a partnership activity between the community and the observatory. Observatory standards and requirements are met not just by throwing a bunch of documents over the wall to the community groups, but by developing active, close working relationships that ensure that the delivered hardware and software meets standards, meets requirements, and performs as required scientifically. This surely is what will be needed within UC. It will be a challenge for stand-alone ad-hoc instrument groups to routinely have the breadth of experience to deliver an instrument to the requirements of TMT or Keck without a close working relationship with UCO. The TMT and Keck instrument support groups will certainly provide support where they can, but they will have to deal with instrument development across the full partnership and will likely be stretched thin (as is the case at Keck currently, and is likely to be the case for TMT). There is substantial benefit in having experienced personnel at UCO who can help any new UC instrument PI and their team ensure that the UC instruments meet the goals and needs of the UC astronomy community.

*(e) facility operations.* Operations of facilities will likely be a fluctuating aspect of UCO's future, scaling back at Mt Hamilton because of the budget pressures, but offset by the probable need to carry out operations in the future as new opportunities are developed. The role that this aspect will play longer-term is TBD, but it does require a somewhat different skill mix and people mix than the other aspects.

The loss of these activities would have a major impact on UC astronomy and greatly reduce, over time, UC's standing as an leading international player in astronomy. Given the scale of the Moore Foundation gift and the visibility that UC has accrued from the success of its astronomy program, I hope that ways are found to move forward and that UC astronomy is positioned to deal with the above activities (but see § 6 below).

Having the right people of staff is key to carrying out these activities. This begins with the astronomy faculty. The UCO faculty are akin to the 50-50 astronomers/physicists at other observatories who routinely provide much of the leadership. A committed, experienced group of astronomers is crucial if the Observatory is to be recognized as a forefront entity and to function effectively in the numerous situations where UCO represents UC astronomy (projects, private funding, dealings with funding agencies, budget reviews, etc.). The optimal number of such people is part of the discussions about the future of UCO, but the importance of having a significant core group of highly-regarded astronomers as the recognized leaders of UCO cannot be underestimated.

To this it is crucial to add engineers and managers who elicit comparable respect for their abilities and achievements. I cannot emphasize enough the importance of having a strong, experienced and highly credible team of scientists (astronomers/physicists), engineers and managers at the heart of the observatory.

At this point it is worthwhile to be reminded of the goals and priorities that the Astronomy Task Force (ATF) identified. The ATF recommendations were clear and well-stated and are reproduced directly here:

## *Prioritized Investment Recommendations*

1. ***Ensure the long-term success of UC leadership within the TMT project.*** *UC should continue to play a leadership role in the development of TMT's telescope design and instrument suite by investing in the technical expertise and UC laboratories. UC should commit to shifting $6.5 M/yr in 2018 from Keck operations to TMT operations when Caltech is contractually obliged to pick up that portion of Keck operations. This represents UC's contribution to TMT operations for a 15 – 18% share, leaving UC's share in Keck unchanged.*
2. ***Keck the Keck Observatory at the cutting-edge of 10-m class telescopes and maintain UC's current share of the telescopes.*** *UC should continue the contractually obliged funding of Keck operations. It should design and construct new instruments and new adaptive optics systems for the Keck Observatory. This requires UC to keep its instrumentation labs strong (at UCSC and UCLA) and to pursue, with its Keck partners, sources of additional funding.*

3. ***Strengthen support for development and construction of instrumentation and adaptive optics.*** *UC facilities, instruments, and personnel are vital to UC's leadership in both Keck and TMT and to the success of these observatories. UC should focus system-wide funding on labs capable of building next generation AO and instrumentation. It should also identify ways to mitigate risk for TMT and advance science at Keck.*
4. ***Continue funding Lick Observatory at current levels, while exploring new funding models.***

**The bottom line: To meet these goals UCO is essential, with the right mix of committed and highly experienced astronomers, engineers and managers.**

**UCO is not just about building instruments.**

## (6) Does UCO have a future?

As I noted above dealing with the budget issues can be accomplished if the right framework is adopted:

1. A rational process of review and assessment by those funding the program with the goal of optimizing the program;
2. An understanding of how observatories work and what makes them successful;
3. A willingness to step beyond parochial self-interest to find compromise solutions that are optimal for the overall program;
4. A willingness to work with observatory people who are trying to develop options for the future.

The budget pressures and the priorities identified by the Astronomy Task Force and the External Review Committee make it necessary to work at both revising the skill mix and optimizing efficiency and productivity of UCO. However, the resulting changes to the organization need to be done thoughtfully, utilizing an understanding of what makes an observatory function well, as UCO is recognized to have done, and done with care to maximize the future potential for UC astronomy, while minimizing the damage.

Is this happening?

Unfortunately, a number of events suggest that the discussion regarding UCO's future is not being carried out in a thoughtful, objective way. Five indications of this are noted here.

**First, astronomy is not taking a disproportionate level of UC resources.** The belief that astronomy takes a disproportionate level of UC resources has become rather pervasive. However, the "evidence" being used to demonstrate this from the Office of Research and Graduate Studies (OGRS) is incomplete and incorrect. A more complete analysis of the resources expended in sciences across the UC system shows that astronomers receive a typical level of UC resources, less than some areas of science and more than others. The chart below shows the actual situation. This is taken from an extended discussion of research costs and investments in the forthcoming *Cost of Astronomy* report. Unfortunately, the incorrect "evidence" is being used, inappropriately, in an effort to cut funding for UC astronomy. This is not conducive to carrying out a thoughtful and balanced discussion of how to most effectively respond to the current budget challenges.

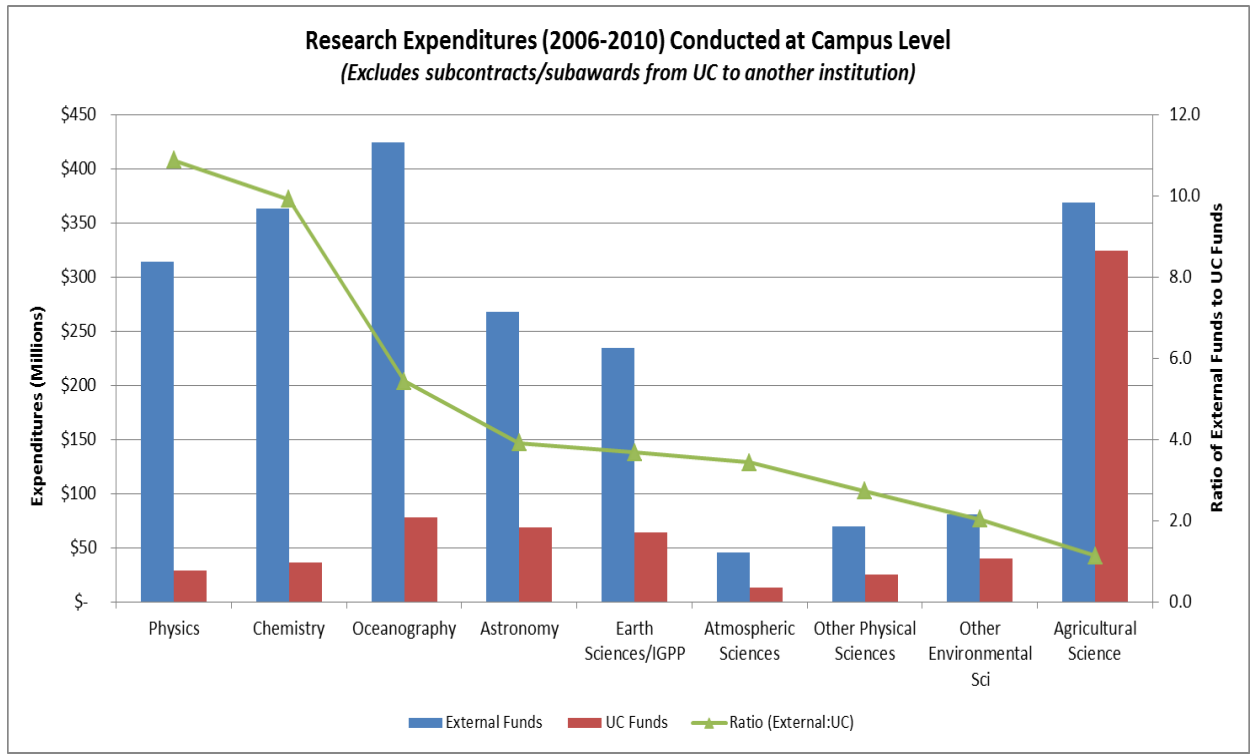

**Second, the new MRU model is not appropriate for UCO.** The concept of the classic "Multi-campus Research Unit" (MRU) in the University of California seems to be evolving in a new direction to become increasingly short-term and repeatedly re-competed. UCO is currently formally classed as an MRU, but efforts to continually force it into the new box are inconsistent with the large scale and multi-decade lifetime of astronomy facilities (telescopes) and with the decade-long timescale for the development and implementation of new instruments and upgrades. UCO needs to be defined, evaluated, reviewed and funded consistent with its long-lived cross-campus nature. This is not arguing for "special" protected status for UCO, but for a rational oversight structure consistent with the nature of UCO's operations and the lifetime of its facilities.

**Third, the $2.6M budget debt is not just the fault of UCO and needs to be fairly assessed.** The situation whereby UCO developed a very large budget debt as a result of misunderstandings, miscommunications, disregard for a long-standing Memorandum of Understanding, and lack of attention to issues over a three-year period has resulted in UCO's being forced to take a very large cut in budget ($2.6M). The fault for this has never been fairly assessed by an independent evaluation. Yet 100% of the fault is attributed to UCO and none to UCOP management, and UCO is being told that it will have to pay back the entire $2.6M debt. This is having a major effect on UCO's ability to plan for the future and respond to present needs for Keck, future needs for TMT, development of new technologies and current efforts at Mt Hamilton.

**Fourth, eliminating UCO faculty positions will decimate UCO.** The recent move to transfer all UCO faculty positions to UCSC and eliminate UCOP funding whenever retirements occur will – if continued indefinitely – leave *NO* faculty associated with UCO except those paid for by UCSC. Since UCSC has no responsibility to support system-wide activities directly from its budget, this will ultimately lead to the loss of any system-wide faculty support. As UCO faculty retire they will not be replaced. *This will decimate UCO within a few years*, will damage the experience base and breadth of capabilities and skills, and render UCO unable to attract needed new people. It will likely also have the effect that the good people will leave when they see that UCO is being eliminated. As noted above, without a strong faculty UCO will not be recognized as a functioning observatory.

**Fifth, the budget uncertainty makes planning very difficult.** UCO has been cutting staff and eliminating positions through retirements for many years (with a layoff as recent as December 2012). UCO has been responding to the budget pressures, as others have done in UC. The need for thoughtful discussions on a budget plan has become critical.

These issues, and others, pose a major challenge for the future of UCO and for the future of UC astronomy. Recent actions presage that astronomy at UC will not be the leading force that it has been in prior decades. However, the situation has not progressed to the point where it cannot be recovered. The future for UC astronomy could be very positive, even with a reduced budget, if we collectively work to optimize the future program through a concerted effort across the system by UC astronomers and with senior management at UCOP. I hope that the process by which decisions are made becomes more transparent, more involving of all the stakeholders, with greater recognition of the role that UCO plays, and is done cognizant of the realities and challenges of optimizing existing and planned facilities.

## Acknowledgements:

I greatly appreciate many thoughtful and continuing discussions on the role of observatories with numerous colleagues over the years – with members of the astronomical community in academic departments, in observatory-like centers and in the science agencies. The environments within which observatories function and are funded vary greatly, yet there is surprising convergence internationally on the structure, nature and staffing of such observatories.

## Appendices:

    A. Author's experience summary;
    B. Astronomy Task Force executive summary;
    C. External Review Committee executive summary.

## A. Author's experience summary:

This analysis of observatory characteristics and their roles is based on my extensive experience with observatories and major projects, particularly with oversight of the roles and activities of observatories.

My 40+ years as an astronomer have been in observatories of several different types (national ground - NOAO; space - STScI; academia - UCO). This has included efforts ranging from conceptual development to major roles in ground instrumentation (e.g., DEIMOS concept development; NOAO instruments) and space instrumentation (Advanced Camera (ACS); Infrared (IR) channel for HST's Wide Field Camera 3 (WFC3)), and new facilities (HST; NGST, now

JWST; Keck). My commitment to UC astronomy was focused on the huge effort in the 1990s on Keck, along with many other UCO people. This involved leading committees (SSC; Segment Acceptance Committee), interfacing with the CARA Board, initial development on DEIMOS. There were numerous activities that were crucial to the success of the observatory as an observatory (particularly the Keck SSC which managed the instrument budget). I was chair of the TMT Science Advisory Committee also during a crucial period where the SAC had to convince the Board to retain the three first light instruments. In addition to my efforts for UCO, the concept of NGST/JWST was developed by me in the late 1980s and early 1990s with Pete Stockman and a very innovative engineer Pierre-Yves Bely. My activities for JWST have included key roles in recent efforts to get JWST back on track (the Independent Comprehensive Review Panel (ICRP), set up by the NASA Administrator at the request of Senator Mikulski) and to keep JWST alive (after the cancellation of JWST by Congress in 2011). My experience has involved management (Deputy-Director at STScI), management oversight (Board Chair for STScI in the late 1990s and early 2000s; AURA Board), Chair of the UV-Optical in Space panel in the 1990 Decadal Survey, Chair of the Astronomy and Astrophysics Advisory Committee (AAAC), a very influential national committee, Chair of the European Southern Observatory (ESO) Visiting Committee, Chair of a key JWST committee (James Webb Space Telescope Advisory Council (JSTAC) and Co-Chair of the JWST Independent External Team.

# 1 Executive Summary

The University of California (UC) Astronomy Task Force (ATF) was established to lead a community process culminating in a set of priorities for future system-wide investment in Astronomy & Astrophysics (A&A). Each UC campus with an A&A program (all but Merced & San Francisco), as well as the UC-run national labs had representation on the ATF. To broadly canvass the UC community for input, the ATF created a web-based Survey, held two town-hall meetings, requested written comments, and held multiple campus-level faculty meetings. Additional information was obtained from presentations from key administrative individuals.

**Key Findings**

UC has a vibrant and world-class A&A program, overseeing arguably the most prolific ground-based telescopes in the world. Over the last two decades, UC members have driven many of the most profound and fundamental contributions of our time, including discoveries of the accelerating universe, planetary systems around Sun-like stars, the anisotropy of the cosmic microwave background, and the supermassive black hole at the Galactic Center. Members of the UC A&A community have also made major technological breakthroughs, most notably, the segmented mirror design of the Keck telescope, which has been widely emulated, and the development of key elements of the Keck laser guide star adaptive optics system. These achievements, and many others, have led to significant rewritings of textbooks, have been featured as "Science Result of the Year" by science-wide publications, and have led to numerous awards and honors for UC faculty. These success stories are frequently the highlights of documentaries shown around the world, providing great visibility for UC's leadership and excellence as well as playing a significant role in the science education of the general public.

Currently, UC is operating shared telescopes at the Keck and Lick Observatories, and it is these shared facilities which have played the central role in UC's preeminence in A&A and the growth of departments across the system. Access to these outstanding facilities has attracted and retained top faculty and students to both large and small UC campuses. Significantly, there has been beneficial impact for those who directly use such facilities as well as for those whose work indirectly profits from the synergies and intellectual vibrancy that these facilities bring. Retention of technological expertise, collaborations with the UC-run national labs, and system-wide investment in a large multi-campus research unit (UCO) for instrument development and centralized operations, has promoted a highly engaged and innovative astronomical community.

For more than a decade, UC has played a central and leading role in the Thirty Meter Telescope (TMT) project, now in the early-construction phase. Technologically, TMT gains major heritage from the Keck Observatory in both telescope design and instrumentation, with UC faculty serving as the Project Scientist (Jerry Nelson, UCSC) and Principal Investigators of two of the three first-light instruments (James Larkin, UCLA; Rebecca Bernstein UCSC). Politically, UCSB Chancellor Henry Yang has served as chair of the TMT Board for the past four years, working tirelessly on critical issues such as the site and the partnerships. Financially, UC has received major philanthropic gifts for the TMT from the Moore Foundation. These gifts are notable within UC both for their size to a single scientific project and their support of a UC system-wide activity, as opposed to a campus-specific one.

## ATF Report Executive Summary

Formally, the UC investments in A&A are separated into three funding lines due to differing partnerships for UCO, Keck Observatory and TMT, but these investments are highly coupled and leveraged off one another in a successful and productive way.

With its large ground-based optical and infrared telescopes, UC will leverage future, high-priority national observatories by enabling key follow-up observations such as faint-object spectroscopy or high angular resolution imaging. Similar to the Keck Observatory's complementary role to the Hubble Space Telescope over the past 15 years, the TMT complements planned national facilities including ALMA, LSST and JWST.

The survey of the UC A&A community clearly identifies the following prioritized ranking of facilities for UC system-wide investment (and associated percentage support):

1. The Thirty Meter Telescope (TMT) Project (90%)
2. Keck Observatory (89%)
3. UC Instrumentation Labs (70%)
4. Lick Observatory (40%)

Other proposed investments included LSST, the Large Synoptic Survey Telescope (20%), a system-wide facility for astrophysical computations (16%), and a radio astronomy facility (10%).

**Prioritized Investment Recommendations**

*1. Ensure the long-term success of UC leadership within the TMT project.* UC should continue to play a leadership role in the development of TMT's telescope design and instrument suite by investing in the technical expertise and UC laboratories. UC should commit to shifting $6.5 M/yr in 2018 from Keck operations to TMT operations when Caltech is contractually obliged to pick up that portion of Keck operations. This represents UC's contribution to TMT operations for a 15—18% share, leaving UC's share in Keck unchanged.

*2. Keep the Keck Observatory at the cutting-edge of 10-m class telescopes and maintain UC's current share of the telescopes.* UC should continue the contractually obliged funding of Keck operations. It should design and construct new instruments and new adaptive optics systems for the Keck Observatory. This requires UC to keep its instrumentation labs strong (at UCSC and UCLA) and to pursue, with its Keck partners, sources of additional funding.

*3. Strengthen support for development and construction of instrumentation and adaptive optics.* UC facilities, instruments, and personnel are vital to UC's leadership in both Keck and TMT and to the success of these observatories. UC should focus system-wide funding on labs capable of building next generation AO and instrumentation. It should also identify ways to mitigate risk for TMT and advance science at Keck.

*4. Continue funding Lick Observatory at current levels, while exploring new funding models.*

*In addition to the facilities above, we recommend creating a UC Astronomy and Astrophysics Council.* This new body will improve the UC A&A community's ability to examine, optimize, and advocate for, the system-wide investments that UC makes in this field.

**System-wide Review of the University of California Observatories**

**Executive Summary**

1. By all criteria the performance of UCO as an organization that supports and advances observational astronomy within the entire UC system has been excellent. Objective evidence for this excellence includes:

   - UC leadership in astronomy through observations on the Keck and Lick telescopes has produced some of the most important astronomical discoveries of the past 15 years, including ground-breaking work in exoplanets, cosmology and black holes;

   - The assembly of what arguably is the leading ground-based optical astronomical instrumentation group in the world. This group provided much of the intellectual impetus for the Keck telescopes, world leadership in developing and implementing astronomical adaptive optics, and leadership of five instruments for the Keck telescopes, and five instruments for the Lick telescopes. They are now leading the design of two first-light instruments for the Thirty Meter Telescope (TMT);

   - Overall productivity in publications and impact of those publications which ranks in the top tier of major astronomical observatories worldwide;

   - The recruitment of outstanding junior and senior faculty to UC, and the expansion of astrophysics on several campuses, making astronomy one of the most visible and high-impact programs in UC.

2. UCO has been a very effective organization in managing the shared facilities and technical resources for the UC campuses. Access to telescope time is managed in a manner which involves all of the stakeholders and balances the principle of access for all with a strongly merit-based peer reviewed time allocation system. It has forged effective working relations with partner organizations for the Keck Observatory and the TMT project at the technical, scientific, and managerial levels.

3. This success in fulfilling its mission, the breadth of this mission (serving eight UC campuses), the large capital investments in its managed facilities, and its international leadership in astronomical instrumentation all strongly justify the continuation of UCO as a multi-campus research unit.

4. The committee broadly endorses the future vision for UC optical and infrared (OIR) astronomy presented by UCO and the UC Astronomy Task Force, which is built around participation in the Thirty-Meter Telescope (TMT) project with continued participation in the Keck Observatory, with operations costs of TMT largely covered by the 50% reduction in Keck obligations after 2018. In offering this endorsement we caution that it would be prudent for UCO and the University to consider carefully the long-term obligations that entry into the TMT project will entail, and the likely impacts that it will have on the infrastructure and staffing of UCO.

5. Fabrication of instruments for TMT will require significant upgrades to the Santa Cruz laboratory facilities, including a lab with a large interior volume, upgrades in optical measuring capabilities and possibly in the machine shops.  Careful planning will be needed to reach the right mix of using outside vendors and internal expertise to make most effective use of limited funding, especially if TMT funds cannot be used for improvements to the laboratories.  Given the expected need for IR technologies for TMT, the UCLA Infrared Laboratory also needs enhanced levels of support to ensure that the group keeps a core staff between large instrument and detector jobs.

6. UCO has made effective use of the facilities on Mt. Hamilton for a variety of purposes including major surveys of exoplanets, supernovae, and active galactic nuclei, instrument engineering, and education and public outreach.  However the $1.8M currently spent annually on Lick is a significant sum in the face of other funding pressures.  The committee would like to see Lick Observatory continue, but as a largely self-supported enterprise with a strong public function.  We encourage the UCO Director, working with interested astronomers from the other campuses, to seek outside funding sources and implement streamlined operations at Lick if this historic observatory is to continue to contribute.  The long-term future of Lick Observatory should be critically examined as part of a strategic planning exercise.

7. The committee is convinced that the presence of a core staff of UCO research faculty instrumentalists in stable appointments is a key element in UCO's success.  Most of these faculty reside in UCSC for critical mass and the efficient use of technical resources.  This proven model should not be dismantled.  We did not fully examine the rationale for maintaining the current number (14) of "80/20" positions, but are concerned that eventually the cost of maintaining this level of staffing will compete with other UCO priorities.  This issue should be addressed as part of our recommended strategic planning process, in the context of future needs in the TMT+Keck era.

8. Given the importance of UCO's role and the considerable resource it manages it is important to improve and strengthen its system of governance.  We recommend the establishment of a board, largely composed of members external to UC, and including a representative of the UC Academic Senate, who would serve as trustees and advocates for the UCO program.  The main roles of this UCO Board would be to give support and advice to the UCO Director on important policy and management matters; review and approve annual program plans and longer-range strategic plans; evaluate progress against those plans; periodically review the performance of UCO and its Director, and recommend the appointment of a new Director when a vacancy arises.  It should report annually on its activities to the UCOP.

9. We recommend that the UCO Advisory Committee (UCOAC) be retained as the primary conduit for engaging the UC astronomical community in the management of the Observatories.  The role of the committee, however, should be expanded to include discussions of policies, priorities, and plans, with a more formal structure of feedback and response between the UCOAC and the Director. We envisage that the UCOAC would report jointly to the Director and to the Board described above.